\documentclass[aps,prd,twocolumn,showpacs,groupedaddress,nofootinbib]{revtex4}
\usepackage{graphicx}
\usepackage{color}
\begin{document}

\title{Search for H dibaryon on the lattice}

\author{Zhi-Huan Luo$^{1,2}$, Mushtaq Loan$^{3}$ and Yan Liu$^{2}$ \\
$^{1}$ School of Physics and Engineering, Sun Yet-Sen University, Guangzhou 510275, P. R. China \\
$^{2}$ Department of Applied Physics, South China Agricultural University, Guangzhou 510642, P. R. China \\
$^{3}$ International School, Jinan University, Guangzhou 510632, P.
R. China}

\date{\today}

\begin{abstract}
We investigate the H-dibaryon, an $I(J^{P})=0(0^{+})$ with $s=-2$,
in the chiral and continuum regimes on anisotropic lattices in
quenched QCD. Simulations are performed on modest lattices with
refined techniques to obtain results with  high accuracy over a
spatial lattice spacing in the range of $a_{s} \sim 0.19 - 0.40$
fm. We present results for the energy difference between the
ground state energy of the hexa-quark stranglet and the free
two-baryon state from our ensembles. A negative energy shift
observed in the chirally extrapolated results leads to the
conclusion that the measured hexa-quark state is bound. This is
further confirmed by the attractive interaction in the continuum
limit with the observed H-dibaryon bound by $47 \pm 37$MeV.

\end{abstract}
\pacs{11.15.Ha, 14.40.Lb, 12.38.Gc, 12.38.Mh}

\maketitle

\section{Introduction}

Search for dibaryons is one of the most challenging theoretical
and experimental problems in the physics of strong interaction. In
the non-strange sector, only one dibaryon, the deuteron, is known
experimentally. In the strange sector, on the other hand, it is
still unclear whether there are bound dibaryons or dibaryon
resonances. Among others, the flavor-singlet state (uuddss), the
H-dibaryon, has been suggested to be the most promising candidate
\cite{Farhi-1984-PRD}. The H-dibaryon may also be a doorway to
strange matter that could exist in the core of neutron stars and
to exotic hyper-nuclei \cite{Greiner-1998-arxiv,Alcock-1991-NP}. A
deeply bound H with the binding energy $B_{H} > 7$ MeV from the
$\Lambda\Lambda$ threshold has been ruled out by the discovery of
the double $\Lambda$ nuclei \cite{Takahashi-2001-PRL}. The
double-hypernucleus events have either more than one
interpretation for the products or the possibility of production
of excited states. The analysis of the events in
\cite{Takahashi-2001-PRL} ignored the possibility that the single
or double hypernucleus was produced in an excited state in which
case the value of the binding energy would increase by the
excitation energy.  Thus there still remains a possibility of a
shallow bound state or a resonance in this channel. Since the Bag
Model prediction of a deeply bound H-dibaryon,  with large binding
energy of $O(100)$ MeV \cite{Jaffe-1977-PRL}, many experiments
have been triggered to look for this possible particle, but few of
them have confirmed the existence of the H-dibaryon
\cite{Shahbazian-1990-PL,Shahbazian-1993-PL,Yemelyanenko-1999-PB,Trattner-2006-PhD,Yoon-2007-PRC}.
The experiments were inspired by the Skyrme model prediction, and
the experimental discoveries have in turn spawned intense interest
on the theoretical side, with studies ranging from chiral soliton
and large $N_{c}$ models, quark models, phase-shift analysis and
QCD sum rules \cite{Sakai-2000-PTPS}. Summarizing the previous
theoretical and experimental investigations a slightly bound or
unbound H-dibaryon is predicted.

In the search for such exotic states lattice QCD plays an
important role in which precise predictions for hadronic
observables with quantifiable uncertainties are made.  A
considerable interest in H-dibaryon started with Jaffe's work
\cite{Jaffe-1977-PRL} demonstrating the role of the chromomagnetic
interaction in the stability of light multiquarks. Since then a
number of quenched LQCD calculations have been performed for the
search of the H-dibaryon but no definite conclusions have been
reported. Earlier lattice investigations
\cite{Mackenzie-1985-PRL,Iwasaki-1988-PRL} gave somewhat mixed and
contradicting results on the spatially localized resonance status
of H-dibaryon. These studies however, suffered from low
statistics, relatively large quark masses and considerable
finite-size effects could not be rules out for the smaller lattice
size. More precise studies on large volumes concluded an unbound
H-dibaryon in infinite volume limit
\cite{Negele-1999-NPB,Wetzorke-2000-NPB,Wetzorke-2003-NPB} while
others reported hints of a bound H-dibaryon for a range of
light-quark masses \cite{Luo-2007-MPLA}.

Very recently, NPLQCD and HALQCD Collaborations reported results
from their fully dynamical lattice calculation that shed new light
on the status of H-dibaryon \cite{Beane-2011-PRL,HAL-2011-PRL}.
NPLQCD Collaboration presented a strong evidence for a bound
H-dibaryon from their calculations performed in four lattice
volumes. Using Lusher method \cite{Beane-2004-PRB} to extract
two-particle scattering amplitude below threshold, NPLQCD
collaboration found H-dibaryon bound by $16.6$ MeV at a pion mass
of $m_{\pi} = 390$ in the infinite volume limit
\cite{Beane-2011-PRL}. Using their recently proposed approach of
baryon-baryon potential \cite{Ishii-2007-PRL}, the HALQCD
Collaboration performed calculations in three lattice volumes at
three different quark masses  and reported a bound H-dibaryon with
the binding energy of $30 - 40$ Mev for the pion mass of $673 -
1015$ MeV \cite{HAL-2011-PRL}. These calculations provide strong
evidence of capability of Lattice QCD in calculating the energy of
simple nuclei, with H-dibaryon being an example. Having said that,
it still remains an open question whether the H-dibaryon is bound
at the physical point and with the inclusion of the electroweak
interactions.  This provides a strong motivation for pursuing
numerical calculations at smaller lattice spacings, and over a
range of quark masses including those of nature.

The target of this work is to address the status of the H-dibaryon
by calculating the mass differences between the candidate
H-dibaryon and the free two-baryon states in the continuum limit
and at physical quark masses. Using some refined methods and
techniques, we carry a multi-lattice spacing analysis at and near
physical pion mass on improved anisotropic lattices with an
attractive feature that with modest lattice sizes one can access
large spatial volumes while having a fine temporal resolution.
Rather than extracting the hadron mass from the ratio of two
temporal nearest correlators, the Levenberg-Marquardt algorithm is
adopted to solve the hyperbolic-cosine ansatz of hadron
correlation functions. This is very useful in finding a larger
temporal fit range, hence more clear signals for precise hadron
masses. Continuum limit is also considered in this work which will
provides  a real physical picture of the H-dibaryon.

The rest of the paper is organized as follows. The technical
details of the lattice simulations are discussed in  Sec. II,
where we outline the construction of the correlation functions
from interpolating operators and the actions used in this study.
This section also discusses the procedures of chiral and continuum
extrapolations. The results are presented and discussed in Sec.
III, where we attempt to take the chiral and continuum limits and
address chiral, finite-spacing and quenching effects. This sets
the stage for a discussion of lattice resonance signature of the
H-dibaryon lying lower than two-$\Lambda$ channel masses in the
physical regime. Finally, we present our conclusions in Sec. IV.

\section{Simulation Details}
\subsection{Choice of Interpolating fields}
The explicit construction of the operator for H-dibaryon requires
the symmetrization of the colour and spinor indices of two
triplets of quarks in order to obtain colour and spin singlet. Our
choice for the appropriate operator is  motivated by possible
structure of H-dibaryon and based on the idea of diquark
formulation  and has the following form \cite{Donoghue-1986-PRD,
Golowich-1992-PRD}:
\begin{eqnarray}
O_{H}(x) & = & 3(udsuds)-3(ussudd)-3(dssduu)\nonumber\\
(udsuds)
 & = & 3\epsilon^{abc}\epsilon^{def}(C\gamma_{5})_{\alpha\beta}
 (C\gamma_{5})_{\gamma\delta}(C\gamma_{5})_{\epsilon\phi}
 \nonumber\\
 & & \times
 \left[u^{a}_{\alpha}d^{b}_{\beta}s^{c}_{\epsilon}u^{d}_{\gamma}d^{e}_{\delta}s^{f}_{\phi}\right],
\end{eqnarray}
where the roman letters denote the colour indices, greek letters
represent the spinor indices and $\epsilon^{abc}$ the usual
antisymmetric tensor defined over the range of their indices.
Taking the symmetry properties of the $\epsilon$-tensor and the
$(C\gamma_{5})$-matrix under the interchange of two indices into
account, the H-dibaryon correlation function can be obtained from
\begin{equation}
C_H(\vec{x},t)=<O_H(\vec{x},t)O_H^\dagger(0)>.
\end{equation}
The hadron masses $M_{h}$ and $M_{\Lambda}$, needed to obtain the
energy shift $\Delta E = E_{h}-2m_{\Lambda}$, are calculated by
the fitting the correlation functions with a
multi-hyperbolic-cosine ansatz
\begin{equation}
\label{eqn-cor-fit}
C(t)=\sum_{i=0}^nA_i\cosh[m_i(T/2-t)],
\end{equation}
where $m_i$ is the effective mass of the $i$th excited state and
$A_i$ the amplitude corresponding to this state. The method of
calculation is straightforward in principle, not differing
essentially from the calculation of hadron masses. We determine the
mass of the ground state for each particle, and the mass difference
$\Delta M (H-2\Lambda)$ from the fit. In order to reduce the
contaminations of the excited states, the maximum time separation is
used to extract the results. This is achieved by adopting the
Levenberg-Marquardt algorithm to solve this nonlinear least-squares
fit. The hyperbolic-cosine fits are performed over the time interval
in which an acceptable value of the probability, used to estimate
the goodness-of-fit of the data, is obtained. Considering the
contribution of ground state only, the correlation function is
fitted by the form
\begin{equation}
\label{eqn-cor-fit-ground} C(t)=A_0\cosh[m_0(T/2-t)].
\end{equation}
To account for the strong correlation of data in time, we use the
full covariance matrix to construct the $\chi^2$ function
\begin{equation}
\chi^2=\sum_{i,j}[C(t_i,A_0,m_0)-D(t_i)]M_{ij}^{-1}[C(t_j,A_0,m_0)-D(t_j)],
\end{equation}
and obtain the the covariance matrix $M_{ij}$ as
\begin{equation}
M_{ij}=\frac{1}{N_c(N_c-1)}\sum_{k=1}^{N_c}[D(k,t_i)-D(t_i)][D(k,t_j)-D(t_j)],
\end{equation}
where the $D(k,t_i)$ is the $k$th correlator and $D(t_i)$ the mean
value of the correlator at time $t_i$, $N_c$ denotes the total
number of configurations. At minimum $\chi^2$, the gradient of
$\chi^2$ with respect to the parameters $(A_0,m_0)$ will be zero
and the mass of ground state $m_0$ is estimated.

\subsection{Anisotropic Lattice Actions}
To examine the H-dibaryon in lattice QCD, we explore the improved
actions on anisotropic lattices. These action  display nearly
perfect scaling, thus lattice-spacing artifact contributions are
expected to be small, and providing reliable continuum limit
results at finite lattice spacings can be obtained. With most of
the finite-lattice artifacts having been removed, one can use
coarse lattices with fewer sites and much less computational
effort. Using a tadpole-improved anisotropic gauge action
\cite{Morningstar-1997-PRD}, we generate quenched configurations
on a $12^{3}\times 60$ lattice at five couplings in the range
$\beta = 2.0 - 4.0$ and at a bare anisotropy of $\xi = 5.0$.  We
generated $500$ gauge field configurations for each lattice and
the configurations are separated by $100$ compound sweeps after
skipping $1000$ sweeps for the thermalization. We define  a
compound sweep as $5$ over-relaxation \cite{Creutz-1980-PRD}
sweeps followed by one Cabbibo-Marinari \cite{Cabibbo-1982-PLB}
sweep.

For the fermion fields,  we employ the space-time asymmetric
clover quark action on anisotropic lattice
\cite{Okamoto-2002-PRD,Harada-2002-PRD} with spatial Wilson
parameter $r_{s}=1$. The clover improvement coefficients $c_{s,t}$
are estimated from tree-level tadpole improvement whereas for the
ratio of hopping parameters $\zeta = K_{t}/K_{s}$ we adopt both
the tree-level improved value and a non-perturbative one. Since it
becomes harder to obtain a reasonable signal-to-noise ratio at
lighter quark masses for the multi-quark systems, we employ
relatively heavy quark masses in our calculations. The bare
strange quark mass is set by measuring the $s\bar{s}$ pseudoscalar
mass at four heavy quark hopping parameters $\kappa_{h}$. At each
strange quark mass, hadron propagators are measured for six light
hopping parameters $\kappa_{l}$ such that the mass ratio of
$M_{K}/M_{N}$ compares well with the experimental value. Our
quenched quark propagators cover a range of quark masses,
corresponding to pion masses from $1325$ MeV down to $500$ MeV. We
also considered two smaller masses, but find that the signal for
these becomes highly unstable, hence do not include these in our
analysis.
\subsection{Smearing technique}
To increase the overlap of the operators with the ground state, all
of the hadronic correlators were calculated using the method of
smearing the interpolating operator, essentially making the hadronic
operator spread around their central location in space. In this
study, we use the gaussian smearing  which is obtained by replacing
the quark field $q(x)$ by the smeared quark field $q_{smear}(x)$
defined as \cite{BERNARD-1998-NPB}
\begin{equation}
q_{\rm{smear}}(t,\vec{x})=N\sum_{y}\exp\{-\frac{|\vec{x}-\vec{y}|^2}{2\rho^2}\}q(t,\vec{y}),
\end{equation}
where $N$ is an appropriate normalization factor and $\rho$ the
smearing size parameter. This technique has numerical advantages
since the smearing function separates into two factors one
belonging to the quark and the other to the antiquark, thus will
help to maximize the ground state contribution relative to the
ones of the excited states. The problem is that the smeared
operators are no longer gauge-invariant because the quark and the
antiquark are spatially separated. We employed Coulomb gauge
fixing to overcome this problem.

\subsection{Extrapolation to the physical quark mass and Continuum limits}
Chiral extrapolations of the H-dibaryon mass and binding energy to
the physical point are important issues. In the exact SU(3) flavour
symmetry, the non-interacting $I(J^{P}) = 0(0^{+})$ with strangeness
$s=-2$ ground state is multiple degenerate, comprised of the states
$\Lambda\Lambda, N\Xi$ and $\Xi\Xi$ with H-dibaryon as the ground
state. A tightly bound H-dibaryon would indicate the chiral
expansion of the form of that for single hadrons. The chiral
extrapolation of single hadrons, such as the lowest-lying octet
baryon masses, is an ongoing topic of discussion and motivates a
deeper understanding of extrapolation form. The baryon chiral
perturbation theory seems reluctant to reproduce LQCD results for
the octet baryon masses, including the results for nucleon mass.
Leinweber {\emph {et al} \cite{Leinweber-2004-PRL} demonstrated that
the chiral extrapolation method based upon finite-range regulator
leads to extremely accurate value for the mass of physical nucleon
with systematic errors of less than one percent.

To addressed the challenges of SU(3) chiral perturbation theory to
describe the baryon masses, Walker-Loud {\emph et al}., detailed a
comprehensive chiral extrapolation analysis of the octet and
decuplet baryon masses, using both the continuum SU(3) heavy
baryon $\chi PT$ as well as its mixed action generalization
\cite{Walker-Loud-2008,Walker-2009-PRD}. The results placed
signature of linearity of nucleon mass in $m_{\pi}$, providing a
remarkable agreement with both the lattice data as well as the
physical nucleon mass. This is in contrast with the expectations
of chiral limit expansion of the general form $M_{N}(m_{\pi})=
a+bm_{\pi}^{2}+O(m_{\pi}^{3})$, where $a$ and $b$ are parameters
determined from the lattice QCD data. Sharpe and Labrenz
\cite{Sharpe-1996-PRD} also found a more complicated and available
form of chiral expansion of baryon masses in quenched
approximation. The problem is that we have only several binding
energy for each lattice spacing, using the form made by Sharpe and
Labrenz with so many coefficients may provides an unavailable and
unreliable fit. Considering the form of binding energy is not
known yet, we apply the general one to perform the chiral
extrapolation and in fact we obtain a good result.

To avoid the ambiguity in the chiral limit estimates, we
extrapolate mass difference and $\Delta M= M_{H}-2M_{\Lambda}$ and
mass ratios $\Delta M/M_{\Lambda}$ using the simplest ansatz
consistent with leading order chiral effective theory,
\begin{equation}
f = \alpha +\beta x, \label{fit1}
\end{equation}
where $x$ represents the pion mass squared and $\alpha$ and $\beta$
are fit parameters. The quantities $f$ and $x$ are accompanied by
statistical errors. We intend to find the combination of $\alpha$
and $\beta$ which minimizes
\begin{equation}
\sum_{i}\frac{\big(f(\alpha,\beta ; x_{i})-\langle
f_{i}\rangle\big)^{2}}{\sigma^{2}_{f_{i}}+\beta^{2}\sigma^{2}_{x_{i}}},\label{fit2}
\end{equation}
where $i$ indexes different data points $\{x, f\}$ and $\sigma$ is
the statistical error of each quantity. The extrapolation is taken
to physical point at fixed strange mass and $M_{\Lambda}$ is taken
as experimental input to make physical predictions.

The continuum extrapolation for the chirally extrapolated
quantities is another important issue in lattice calculations. The
possible error that might effect the simulation results comes from
the scaling violation for our action. Expecting that dominant part
of scaling violation is largely eliminated by tadpole-improvement,
we adopt an $a_{s}^{2}$-linear extrapolation to the continuum
limit, since the lattice-spacing artifacts in our calculations are
expected to scale as $O(a_{s}^{2})$. Also, since the
$O(a_{s}^{2})$ effects largely cancel in forming the binding
energy, we expect such contributions to be small.

\section{Results and Discussion}
Typical examples of the effective mass plot at $(\kappa_l,
\beta)=(0.3110, 3.60)$ and $(0.3115, 4.00)$ are shown in Fig.
\ref{fig1}. As can be seen, smearing improves the overlap with the
H-dibaryon ground state resulting in an earlier plateau.
Consequently the contributions of excited states were substantially
reduced. We find clear signals up to larger time separations with
insignificant statistical fluctuation domination. The fit range
$[t_{min},t_{max}]$ is determined by fixing $t_{max}$ and finding a
range of $t_{min}$ where the ground state is stable against
$t_{min}$. The statistical error analysis is performed by a
single-elimination jackknife method and the goodness of the fit is
gauged by the $\chi^{2}$ per degree of freedom, chosen according to
criteria that $\chi^{2}/N_{DF}$ is preferably close to $1.0$.  The
resulting effective masses of H-dibaryon and $\Lambda$ states for
other values of $a_{t}m_{\pi}$ at $\beta=3.60$ are tabulated in
Table \ref{tab1-beta-3.60}.
\begin{figure}[!h]
    \begin{tabular}{cc}
      \resizebox{40mm}{!}{\includegraphics{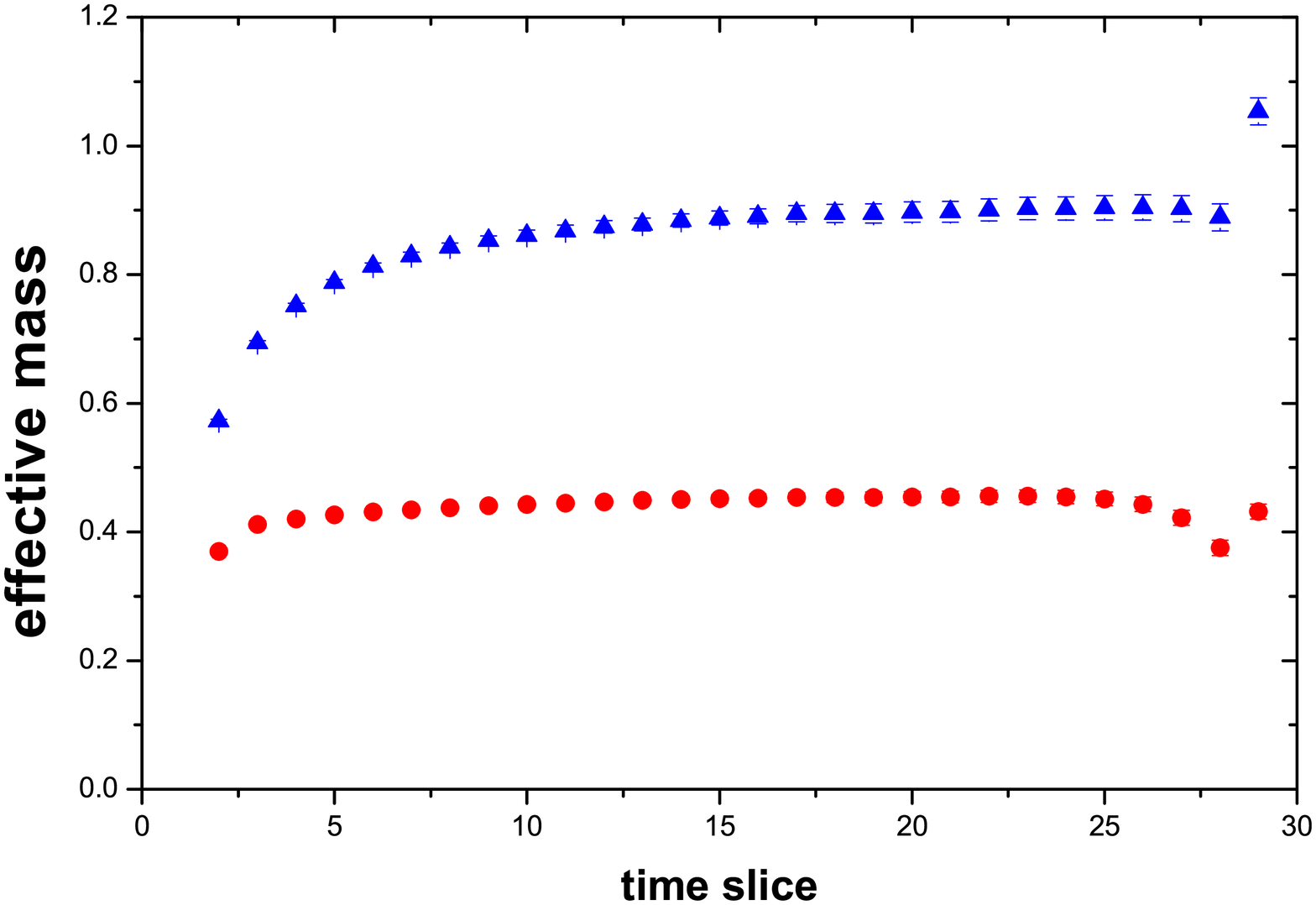}} &
      \resizebox{40mm}{!}{\includegraphics{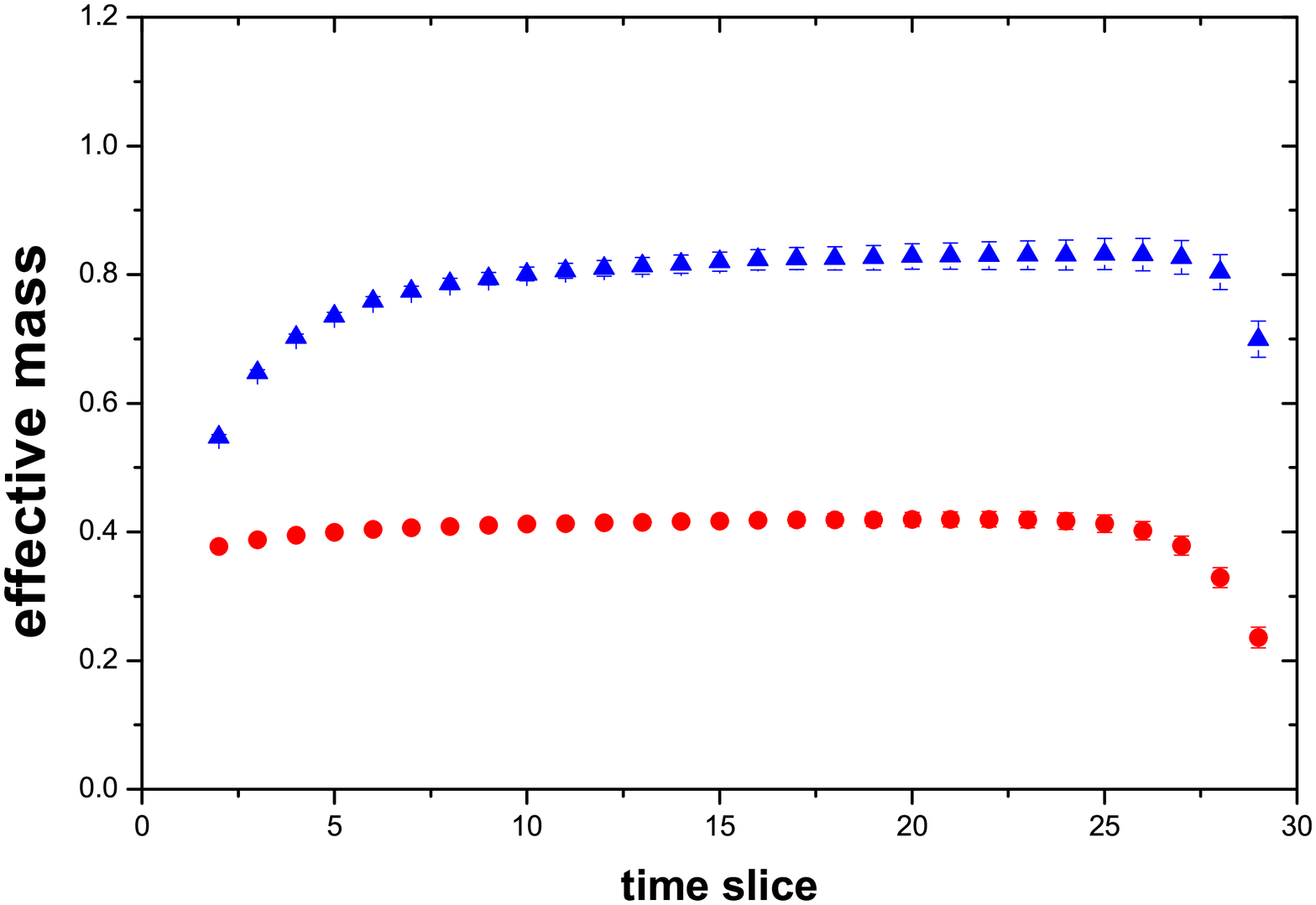}} \\
    \end{tabular}
    \caption{\label{fig1} Effective masses of the $I(J^{P})=0(0^{+})$ stranglet
    (solid triangles) and lambda baryon (solid circles) at $(\kappa_l, \beta) =
    (0.3110, 3.60)$ (left panel) and $(\kappa_l, \beta)=(0.3115, 4.00)$
(right panel).}
\end{figure}
\begin{table}[!h]
\begin{center}
\tabcolsep0.2in \caption{Effective masses of lambda baryon and
H-dibaryon on the $12^3\times60$ lattice at $\beta=3.60$ for various
values of $a_{t}m_{\pi}$.} \label{tab1-beta-3.60}
\begin{tabular}{ccc}\hline\hline
$a_{t}m_{\pi}$  & $a_{t}M_{\Lambda}$ & $a_{t}M_{h}$\\
\hline
0.38847(77) &  0.5202(24) &  1.0320(71) \\
0.36661(78) &  0.4979(24) &  0.9865(71) \\
0.34327(78) &  0.4746(24) &  0.9386(72) \\
0.31759(77) &  0.4489(24) &  0.8868(72) \\
0.28848(77) &  0.4199(24) &  0.8287(73) \\
0.25438(77) &  0.3861(24) &  0.7606(73) \\
\hline\hline
\end{tabular}
\end{center}
\end{table}

In order  to determine the energy difference $\Delta M =
M_{h}-2M_{\Lambda}$ precisely, we work in a regime where $t$ is
small enough that $t\Delta M <<1$, and at the same time $t$ is
large enough that the contributions of excited states are
suppressed. Using the Levenberg-Marquardt algorithm, we indeed
found such a range of $t$ where linear term suffices in the data
presented here. Fig. \ref{fig2} shows the mass splitting between
the H-dibaryon and $2\Lambda$ threshold for the parameter
combination $(0.3110, 0.3115)$ at $a_{s} = 0.211$ and $0.188$ fm,
respectively. In the time interval where a single state dominates,
the plateau region is reasonably consistent with that obtained for
the effective mass of the H-dibaryon. The energy gap shows the
negative value in the plateau region of $12\leq t \leq 22$ and
seems more pronounced with H-mass smaller than two $\Lambda$'s.
The signal of mass difference is dominated by the large
fluctuations in the H-dibaryon correlators beyond $t\simeq 22$.
\begin{figure}[!h]
    \begin{tabular}{cc}
      \resizebox{40mm}{!}{\includegraphics{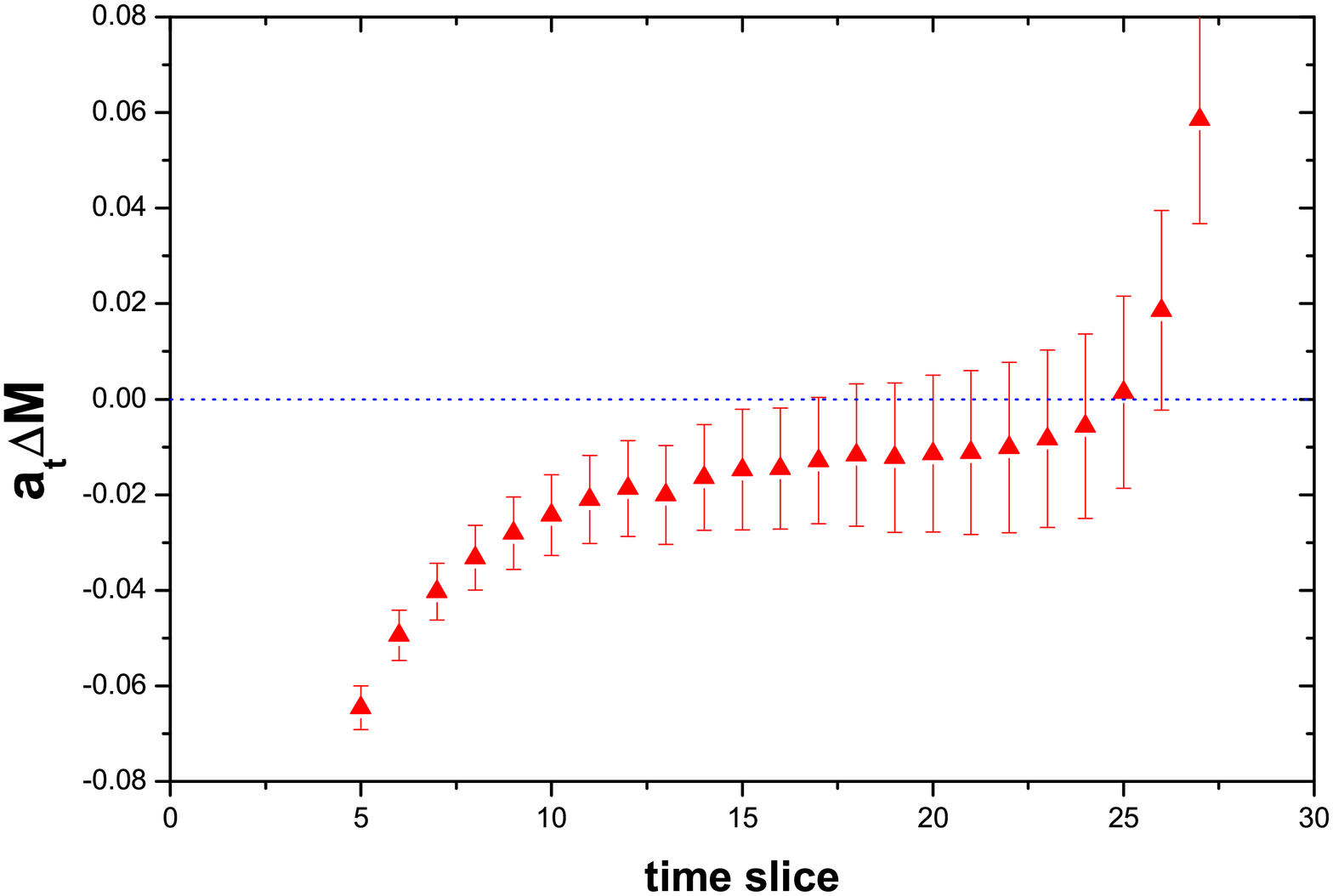}} &
      \resizebox{40mm}{!}{\includegraphics{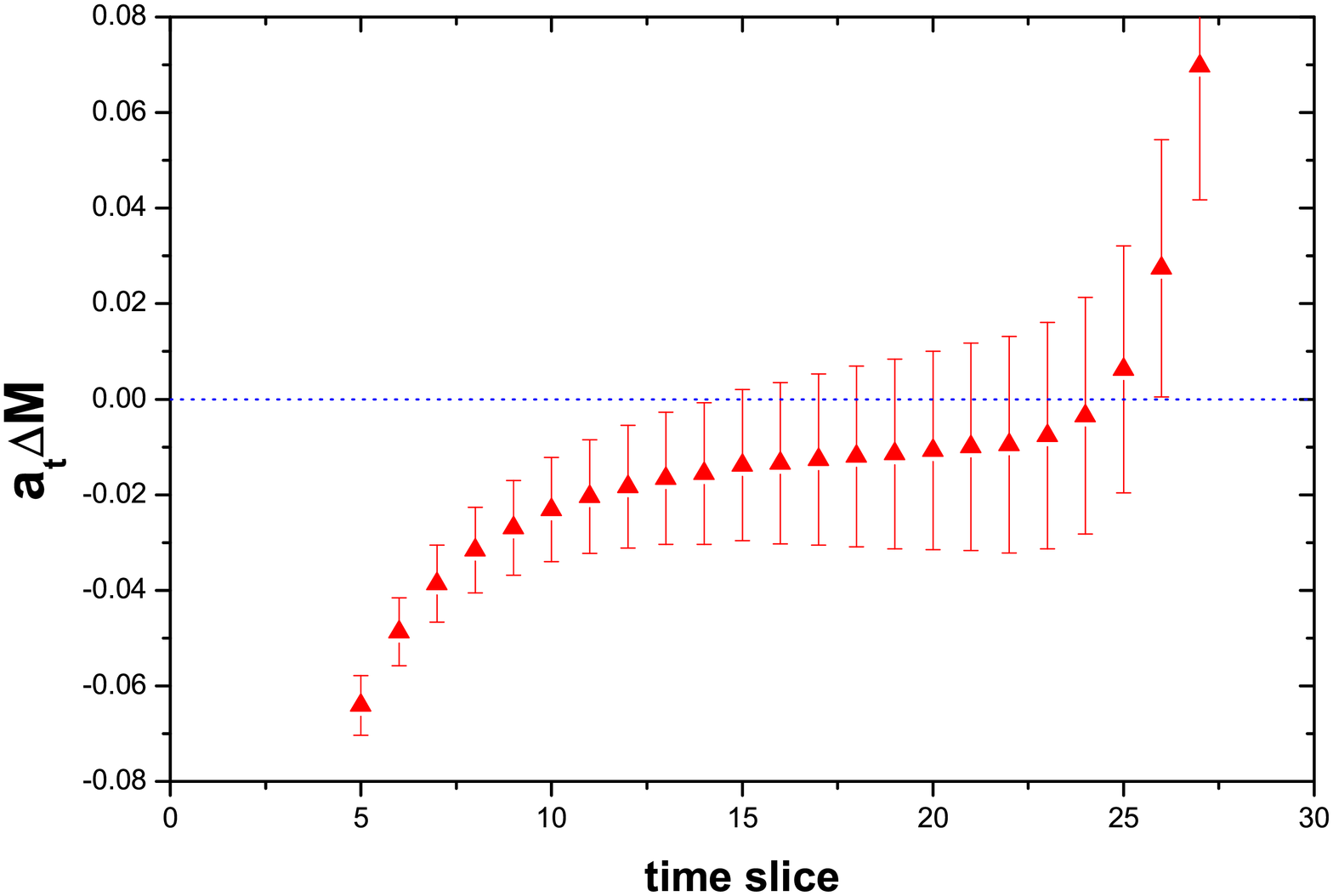}} \\
    \end{tabular}
    \caption{\label{fig2} Effective mass difference between the
H-dibaryon state and the $S$-wave $\Lambda +\Lambda$ two-particle
state at $(\kappa_{l}, \beta )= (0.3110, 3.60)$ (left panel) and
$(\kappa_{l}, \beta )= (0.3115, 4.00)$ (right panel).}
\end{figure}

The results on the other lattice spacings shows consistency in the
behaviour of mass difference over the range of our pion mass range
(see Tables \ref{tab2-beta-2.00}, \ref{tab3-beta-3.20} and
\ref{tab4-beta-4.00}).
\begin{table}[!h]
\begin{center}
\tabcolsep0.2in
\caption{Mass differences and mass ratios between
the H-dibaryon and ($\Lambda +\Lambda$) two-particle state for
various values of $a_{t}m_{\pi}$ at $\beta=2.00$.}
\label{tab2-beta-2.00}
\begin{tabular}{ccc}\hline\hline
$a_{t}m_{\pi}$  & $a_{t}\Delta M$ & $\Delta M/M_{\Lambda}$\\
\hline
0.57298(32) & -0.0351(45) & -0.0396(51) \\
0.55564(31) & -0.0348(42) & -0.0401(49) \\
0.53723(30) & -0.0319(44) & -0.0375(51) \\
0.51770(29) & -0.0335(43) & -0.0402(51) \\
0.49682(29) & -0.0331(42) & -0.0407(52) \\
0.47430(28) & -0.0312(42) & -0.0394(52) \\
\hline\hline
\end{tabular}
\end{center}
\end{table}

With all prerequisites available to measure the energy shift of
H-dibaryon relative to the $2\Lambda$ threshold, we display, in Fig.
\ref{fig3},  the resulting mass differences extrapolated to physical
quark mass value using the ansatz in Eq. (\ref{fit1}). We note that
slope of a linear fit in $m_{\pi}^{2}$ is slightly different at all
lattice spacings. On the other hand, mass difference is almost
constant and weakly dependent on quark mass. Nevertheless the
results on all lattice spacings exhibit a negative value  in the
physical region. The negative mass difference observed in this
region would imply an attractive interaction and hence a signature
of H-dibaryon as a bound state.
\begin{figure}[!h]
\scalebox{0.31}{\includegraphics{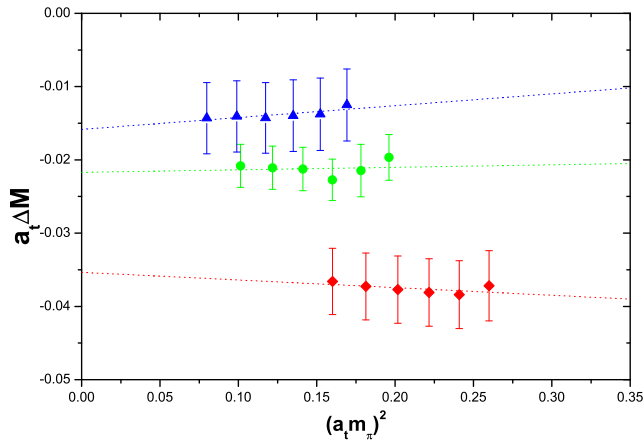}} \caption{
\label{fig3} Effective energy shift between the H-dibaryon state and
$\Lambda +\Lambda$ two-particle state as a function of
$a_{t}m_{\pi}$ squared. Solid circles, diamonds and triangles show
the results at $\beta = 2.00, 2.80$ and $3.20$, respectively.}
\end{figure}

\begin{table}[!h]
\begin{center}
\tabcolsep0.2in \caption{The same as Table \ref{tab2-beta-2.00} but
at $\beta=3.20$.} \label{tab3-beta-3.20}
\begin{tabular}{ccc}\hline\hline
$a_{t}m_{\pi}$ & $a_{t}\Delta M$ &   $\Delta M/M_{\Lambda}$
\\ \hline
0.41147(45) & -0.0125(49) & -0.0223(87) \\
0.39017(46) & -0.0138(49) & -0.0256(92) \\
0.36742(46) & -0.0140(49) & -0.0271(94) \\
0.34252(46) & -0.0143(48) & -0.0291(98) \\
0.31463(46) & -0.0141(49) & -0.0304(104) \\
0.28250(47) & -0.0143(49) & -0.0332(112) \\
\hline\hline
\end{tabular}
\end{center}
\end{table}

Since the quenched spectroscopy is quite reliable for mass ratio of
stable particles, it is physically even more motivating to
extrapolate mass ratio instead of mass. This allows for the
cancellation of systematic errors since the hadron states are
generated from the same gauge configuration and hence systematic
errors are correlated. Fig. \ref{fig4} shows the chiral
extrapolation of the ratio $\Delta M/M_{\Lambda}$ at our smaller
lattice spacings. The ratio shows a weaker dependence on quark mass
and moves into the physical region with a trend of attractive
interaction. In the chiral limit the estimated mass difference at
our smallest and largest lattice spacings are consistent with those
obtained by NPLQCD  and HALQCD Collaborations
\cite{Beane-2011-PRL,HAL-2011-PRL}. The uncertainties from the
chiral fits range from $2$ to $10$ percent for to the smallest to
largest lattice spacings explored here.
\begin{figure}[!h]
\scalebox{0.31}{\includegraphics{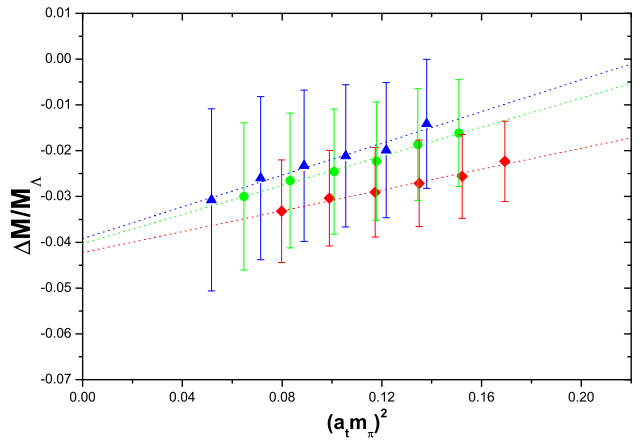}}
\caption{\label{fig4} Plot of mass ratio $\Delta M/M_{\Lambda}$ as
a function of $a_tm_{\pi}$ squared. Solid circles, diamonds and
triangles show the results at $\beta =3.20, 3.60$ and $4.00$,
respectively. Dotted lines are the linear extrapolations to the
chiral limit.}
\end{figure}

\begin{table}[!h]
\begin{center}
\tabcolsep0.2in \caption{The same as Table \ref{tab2-beta-2.00}, but
at $\beta=4.00$.} \label{tab4-beta-4.00}
\begin{tabular}{ccccccccc}\hline\hline
$a_{t}m_{\pi}$ & $a_{t}\Delta M$ &   $\Delta M/M_{\Lambda}$ \\
\hline
0.37144(91) & -0.0070(70) & -0.0141(141) \\
0.34893(91) & -0.0094(70) & -0.0199(148) \\
0.32486(90) & -0.0095(70) & -0.0211(155) \\
0.29802(89) & -0.0098(70) & -0.0233(165) \\
0.26725(88) & -0.0101(70) & -0.0260(178) \\
0.22738(111) & -0.0109(71) & -0.0307(199) \\
\hline\hline
\end{tabular}
\end{center}
\end{table}

The quenching effects might be one of the largest source of the
systematic uncertainties. However, with appropriate definition of
scale, the mass ratios of stable hadrons differ from the
corresponding observed values by less than $6\%$ in quenched
approximation \cite{Butler-1993-PRL,CP-2002-PRD}. In order to
absorb as many quenching effects as possible, we set the scale by
the physical $\kappa_{s}$ by calculating the ratio
$M_{\Lambda}/M_{N}$.  We found that the ratio deviates in the
range of $3 - 4\%$ from its experimental value verifying that the
value of $\kappa_{s}$ used is very close to the physical quark
mass. However, since our calculations use rather heavy pion masses
($> 500$ MeV), the quenching effects are less noticeable within
our statistics. We include a modest estimate of order $5\%$
quenching uncertainties in our analysis.
\begin{figure}[!h]
\scalebox{0.31}{\includegraphics{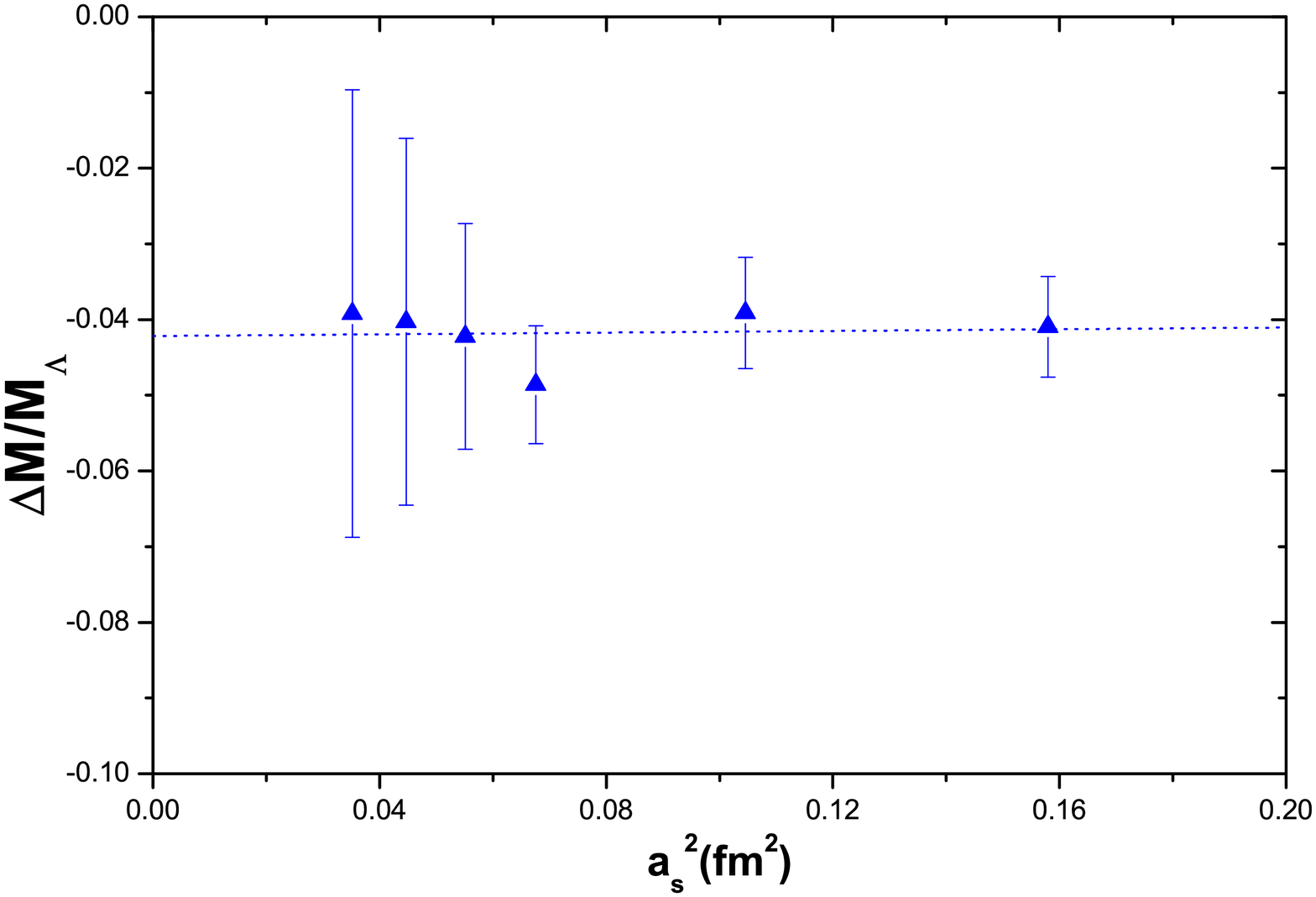}}
\caption{\label{fig5} Compilation of results for the mass ratio,
$\Delta M/M_{\Lambda}$ in the continuum limit. The solid line
represents $a_{s}^{2}$ linear extrapolation to the physical
limit.}
\end{figure}

Whether the energy difference moves in the continuum with an
attractive interaction needs to be explored. Since the
finite-spacing errors in our calculations are expected to scale as
$a_{s}^{2}$, we expect such contribution to have a small effect on
binding energy. Consequently, we expect the observation of
H-dibaryon to survive the continuum extrapolation. We perform the
continuum extrapolation of the chirally extrapolated mass ratios
in Fig. \ref{fig5} and present the results in Table
\ref{tab5-continuum}. Using an $a_{s}^{2}$-linear extrapolation,
we adopt the choice which shows the smoothest scaling behaviour
for the final values, and use other to estimate the systematic
errors.

\begin{table}[!h]
\begin{center}
 \tabcolsep0.3in \caption{Mass ratios between
the H-dibaryon and ($\Lambda +\Lambda$) two-particle state at
various lattice spacings.} \label{tab5-continuum}
\begin{tabular}{cc}\hline\hline
$a_{s}$(fm)  &  $\Delta M/M_{\Lambda}$\\ \hline
0.3974(34)  &  -0.0409(66)\\
0.3234(43)  &  -0.0391(73)\\
0.2599(62)  &  -0.0486(78)\\
0.2347(57)  &  -0.0422(149)\\
0.2114(70)  &  -0.0403(242)\\
0.1875(56)  &  -0.0392(296)\\
\hline\hline
\end{tabular}
\end{center}
\end{table}

As is clear from the figure, the mass ratio shows a weak
dependence on the lattice spacing and varies only slightly over
the fitting range.  Thus we expect our continuum extrapolation
accurate and unambiguous. The continuum extrapolation is
accompanied with an order $8\%$ systematic uncertainty from the
linear fit in $a_s^2$. Using the physical $\Lambda$ mass
$M_{\Lambda} = 1115.68$ MeV, we obtain a continuum estimate of
binding energy $47 \pm 37$ MeV for the binding energy. The
uncertainty shown here results from the statistic and systematic
uncertainties combined in quadrature. The systematic uncertainty
including quenching, chiral and continuum extrapolation effects is
estimated to be of the order of $15\%$. Note that we cannot
estimate the finite size effects since we have been working with
one lattice size. Even though our spatial extent of $L$ is
reasonably large, we cannot rule out the possibility of the volume
dependence of the binding energy in the large volume limit. We
intend to pin down this problem for a conclusive signature in
future work.

\section{Conclusions}

The question of whether the H-dibaryon is a bound or unbound state
is still under debate. The observed negative energy shift pattern
favours the former. In conclusion, we have presented evidence for
existence of bound H-dibaryon state in the physical limit from
quenched lattice QCD calculations. The calculations were performed
over a range of pion masses and lattice spacings using improved
anisotropic lattices with refined computational techniques.
Attractive interaction was found in chiral limit for all pion
masses used in this study and the continuum limit estimate seem to
agree with the predicted value, which is one of the main results
of our paper. Our results seem to be consistent with the recent
results of NPLQCD and HALQCD Collaborations for the  energy shift.
Our analysis takes into account possible artifacts such as,
statistical, chiral and continuum extrapolation uncertainties and
those arising from quenching effects.  On the basis of our lattice
calculation we speculate that the H-dibaryon is to be identified
as bound state. However, the final conclusions will have to await
dynamical simulations incorporating a systematic study of various
possible interpolators that are likely to have a good overlap with
H-dibaryon. We plan to further develop this calculation to involve
a combination of $\Lambda\Lambda - \Xi\Xi -\Sigma\Sigma$
interpolators on larger volumes.

\section{Acknowledgements}

ZHL is grateful to Professor Qiong-Gui Lin for discussions and
valuable suggestions. This work is supported in part by the National
Natural Science Foundation of China (Grant No. 11047021). ML is
supported in part by the Department of Foreign Academic Affairs of
Jinan University. We would like to express our gratitude to the
Theoretical Physics Group at Sun Yat-Sen University for the access
to its computing facility.

\end{document}